\documentclass[twocolumn]{article}
\usepackage{arxiv}
\usepackage{verbatim}
\usepackage{caption}
\usepackage{longtable}
\usepackage{amsmath}
\usepackage{cancel}
\usepackage{bbm}
\usepackage[english]{babel}
\usepackage{authblk}
\usepackage{mathrsfs}
\usepackage{multirow}
\usepackage{tikz}
\usepackage{breqn}
\usepackage{adjustbox}
\usepackage{subfigure}
\usetikzlibrary{quantikz2}

\usepackage{nicefrac}       
\usepackage{microtype}      
\usepackage{graphicx}
\usepackage[square,numbers]{natbib}
\usepackage[colorlinks=true,allcolors=blue]{hyperref}
\title{Digital Quantum Simulations of Hong-Ou-Mandel Interference}
\date{\today}

\author[1]{Navaneeth Krishnan Mohan}
\author[1]{Rikteem Bhowmick}
\author[1]{Devesh Kumar}
\author[2]{Rohit Chaurasiya}

\affil[1]{Qulabs Software (India) Pvt. Ltd}
\affil[2]{Dept. of Physics, Technical University of Munich, Germany}
\affil[ ]{\textit{email: \href{mailto:rikteem.bhowmick@gmail.com}{rikteem.bhowmick@gmail.com}}}

\hypersetup{
pdftitle={Digital Quantum Simulations of Hong-Ou-Mandel Interference},
pdfsubject={Quantum Physics, Quantum Simulations},
pdfauthor={Rikteem Bhowmick,~ Navaneeth Krishnan Mohan,~ Devesh Kumar},
pdfkeywords={Hong-Ou-Mandel Interference,Gray Code},}

\begin{document}

\twocolumn[
  \begin{@twocolumnfalse}
    \maketitle
    \begin{abstract}
      Digital quantum simulation is the process of simulating the dynamics of a physical system by a programmable quantum computer. The universality of quantum computers make it possible to simulate any quantum system, whether fermionic or bosonic. In this work, we discuss the application of digital quantum simulations to simulate a ubiquitous bosonic system, a beam splitter. To perform the boson-to-qubit mapping, we used the gray code, whose superiority over other encoding schemes has been shown recently. We validated our quantum circuit that mimics the action of a beam splitter by simulating the Hong-Ou-Mandel interference experiment. We simulated the experiment in both quantum simulators and actual quantum backends and were able to observe the HOM interference. 
    \end{abstract}  
  \end{@twocolumnfalse}
  ]
\section{Introduction}

Digital quantum simulation (DQS) is an approach introduced by Lloyd \citep{llyod96} to address Feynman's conjecture \citep{feynman1986quantum} regarding the existence of universal quantum simulators. The approach involves using quantum computers to study the dynamics of another quantum system. However, applications of DQS are not only limited to studying the dynamics of the system but involve correlation functions, eigenvalues, partition functions, and so more  \citep{georgescu2014}. Recent progress has been made in this field, and Ref. \cite{bhowmick2023quantum} gives a comprehensive review detailing those advancements across various scientific domains.

DQS can be broken down into three steps: (1) State Initialization, (2) Unitary evolution, and (3) Measurement. The first step is initialising the quantum register to the desired quantum state. Depending on the state to be initialised, it can be complex and efficient algorithms may not exist. In the second step, we decompose the unitary operator that encodes dynamics to local quantum gates. The go-to approach to decompose a unitary operator is trotter decomposition, but alternatives like Linear Combination of Unitaries (LCU) \citep{childs2012hamiltonian} exist. A detailed resource analysis of various Hamiltonian simulation algorithms based on these decomposition methods can be found in Ref. \cite{childs2012}.

At the end of the unitary evolution, we measure the state of qubits to extract the desired information. The complete information about the state is only obtained through quantum state tomography, and it requires measurements and resources that grow exponentially with the number of qubits. Even though the exact reconstruction of the state is costly, there are indirect ways to efficiently calculate quantities of interest, like correlation functions and the eigenvalues of an observable \citep{Somma_2003,francesco_tacchino_digital_2019}. Furthermore, it has been shown recently that measurement overhead can be reduced by using classical shadows (approximate classical description of a quantum state) to measure various physical properties \citep{huang2020}.

A key challenge in quantum simulations is the resource requirement for reliable and efficient execution of quantum algorithms. It largely depends on the mapping scheme to translate the system of interest onto quantum computers. Most gate-based quantum computing architectures follow Pauli's spin-\nicefrac{1}{2} algebra. Thus, spin systems can be directly mapped to quantum computers, and simulation is straightforward \citep {tacchino_spin_2019}. However, mapping is not direct for fermionic and bosonic systems and requires a second quantised representation of the system. In the case of fermions, there exist isomorphic maps that represent the second quantised fermionic states in the language of spins \citep{bhowmick2023quantum}. 


Unlike spins and fermions, which require only finite-dimensional space, bosons inhabit an infinite-dimensional Hilbert space. Thus, simulating bosonic systems necessitates the truncation of the Hilbert space. This truncation modifies the commutation relations of bosons, making it hard to develop isomorphic mapping schemes as in the case of fermions. Nevertheless, one can encode the bosonic states to quantum logical states and then develop one-to-one mapping schemes, as shown in Ref. \cite{Somma_2003}. 

This paper demonstrates how to use DQS to simulate a bosonic quantum system. We have adopted gray code to map bosonic (Fock) states to qubits. There exist various ways to perform boson-to-qubit mapping, and gray code is particularly efficient in terms of fewer qubits and operation requirements \citep{Sawaya_2020}. It has already been demonstrated that Gray code can be used to simulate Deuteron in a quantum computer and found to be better performing than other encoding schemes \citep{Di_Matteo_2021,pooja2021}.

Following the gray code representation, we have mapped bosonic operators to Pauli operators and constructed a quantum circuit that emulates the beam splitter. We validated our circuit by simulating the Hong-Ou-Mandel (HOM) effect/interference experiment on a quantum computer. HOM interference is a two-photon interference that occurs when two indistinguishable photons enter a 1:1 beam splitter, one in each port, as depicted in figure \ref{fig00}. When photons are distinguishable, four potential outcomes are feasible. But if photons are indistinguishable,  two photons will always exit the beam splitter together in the same output mode. A pedagogical review of HOM interference can be found in Ref. \cite{bhongoumandel}.

The article is organised as follows: in section \ref{sec:HOM}, we discuss the theory of HOM interference from a mathematical point of view. In section \ref{sec:QBS}, we discuss the simulation of a beam splitter using gray encoding. This section details how one represents the bosonic creation and annihilation operators in terms of Pauli operators. In section \ref{sec:HOMsimulation}, we discuss the creation of a quantum circuit for simulating the HOM interference and circuit size reduction techniques for real backend simulations. Finally, in sections \ref{results} and \ref{conclusion}, we present the results and provide further discussion on the simulations we performed.

\begin{figure}
    \centering
    \includegraphics[width=0.7\linewidth]{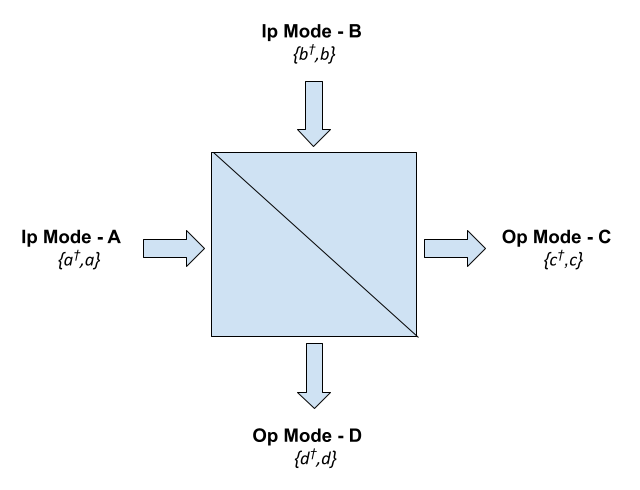}
    \caption{Depicts the input and output modes of the beam splitter. The curly brackets contain creation and annihilation operators of the corresponding mode.}
    \label{figbs}
\end{figure}

\section{Hong-Ou-Mandel Interference} \label{sec:HOM}

\begin{figure}
    \centering
    \includegraphics[width=\linewidth]{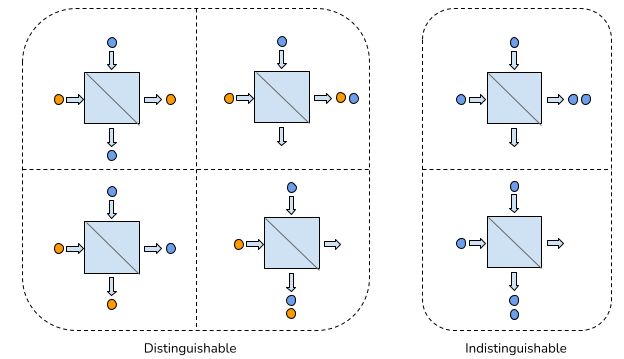}
    \caption{(colour online) Shows the output states when a single photon enters each input mode of the beam splitter. }
    \label{fig00}
\end{figure}
 A beam splitter is a linear optical element which splits the light into a transmitted or reflected beam. HOM interference is a phenomenon with no classical analogue and requires a quantum description to explain it. In the quantum formalism of the beam splitter, we associate bosonic creation and annihilation operators to each mode as depicted in figure \ref{figbs}. The corresponding unitary operator of the beam splitter is given by,
\begin{equation}
    \hat{U}_{B} = \exp(i\theta(b^{\dagger}a + ba^{\dagger})).
    \label{eqn1}
\end{equation}
The angle $\theta = \arctan(R/T)$ where $R$ and $T$ are the reflection and transmission amplitudes of the beamsplitter. 

Consider a 1:1 beam splitter, then we can write the output modes in terms of input modes,
\begin{equation}
\begin{aligned}
    \frac{a-b}{\sqrt{2}}&=c, \ \ \ &\frac{a^{\dagger}-b^{\dagger}}{\sqrt{2}}&=c^\dagger,\\
    \frac{a+b}{\sqrt{2}}&=d, \ \ \ &\frac{a^{\dagger}+b^{\dagger}}{\sqrt{2}}&=d^\dagger.
\end{aligned}
\end{equation}

Let us consider the case of one photon in each input mode A and B ($|11\rangle$). Then, 
\begin{equation}
    \begin{aligned}[b]
        |1\rangle_A |1\rangle_B& = a^\dagger b^\dagger |0\rangle_A |0\rangle_B,\\
        &=\frac{(d^\dagger+c^\dagger)}{\sqrt{2}}\cdot\frac{(d^\dagger-c^\dagger)}{\sqrt{2}} |0\rangle_C |0\rangle_D,\\
        & =\frac{1}{2}\left[\cancel{|1\rangle_C |1\rangle_D} -\sqrt{2} |2\rangle_C |0\rangle_D \right.\\
        & \hspace{2cm} \left. -\cancel{|1\rangle_C |1\rangle_D} +\sqrt{2} |0\rangle_C |2\rangle_D]\right],\\
        &=\frac{1}{\sqrt{2}}\left[|0\rangle_C |2\rangle_D - |2\rangle_C |0\rangle_D\right].
    \end{aligned}
\end{equation}

Hence, one photon in each input mode always comes out in pairs from either of the output modes, as depicted in figure \ref{fig00}. This result is popularly known as Hong-Ou-Mandel interference.
 $$\boxed{|1\rangle_A |1\rangle_B \xrightarrow[]{\hat{U}_B}\frac{1}{\sqrt{2}}\left[ |0\rangle_C |2\rangle_D - |2\rangle_C |0\rangle_D\right]}$$
 
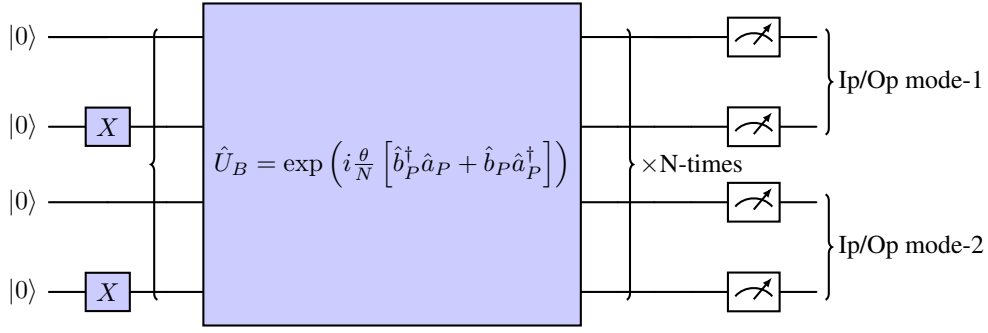
\begin{figure*}[t!]
\centering
\tikzset{
operator/.append style={fill=blue!20},
my label/.append style={above right,xshift=1cm},
phase label/.append style={label position=above}
} 
\begin{adjustbox}{width=0.8\textwidth}
\begin{quantikz} 
    \lstick{\ket{0}}&\qw&\lstick[4]{}&\gate[4]{\hat{U}_{B} = \exp\left(i\frac{\theta}{N}  \left[\hat{b}_{P}^{\dagger}\hat{a}_{P}+\hat{b}_{P}\hat{a}^{\dagger}_{P}\right]\right)}&\rstick[4]{$\times$N-times}& &\qw&\meter{}&\rstick[2]{Ip/Op mode-1}{}\\
    \lstick{\ket{0}}&\gate{X}& & & & & &\meter{}& \\
    \lstick{\ket{0}}& & & & & & &\meter{}&\rstick[2]{Ip/Op mode-2}{}\\
    \lstick{\ket{0}}&\gate{X}& & & & & &\meter{}& \\
\end{quantikz}
\end{adjustbox}
\caption{Quantum Circuit for simulating HOM effect. Decomposition of $\hat{U}_{B}$ requires trotterization, and $N$ represent the number of trotter steps. }
\label{circuit1}
\end{figure*}
\section{Simulation of Beam Splitter}
\label{sec:QBS}

The unitary operator of the beam splitter, as given in equation \eqref{eqn1}, contains creation and annihilation operators, which are infinite-dimensional. However, most quantum computers operate within a finite-dimensional Hilbert space. Therefore, simulating a beam splitter requires truncating the Hilbert space associated with the beam splitter. 

It's crucial to acknowledge that on a finite basis, the creation and annihilation operators no longer adhere to the usual commutation relations $\left[b,a^{\dagger}\right] = \delta_{ab}$. Specifically, the commutation relations becomes
\begin{eqnarray}
\left[b,a^{\dagger}\right]= \delta_{ab}\left[1-\frac{N_{p}+1}{N_{p}!} (b^{\dagger})^{N_{p}}(a)^{N_{p}}\right],
\label{eqn3}
\end{eqnarray}
where $N_{p}$ is the maximum number of photons per mode.

To simulate a beam splitter, we have to find a way to map bosonic creation and annihilation operators to Pauli operators. However, the equation \eqref{eqn3} indicates that the linear span of $b$ and $b^{\dagger}$ is not closed. Consequently, a direct mapping between bosonic and Pauli operators is not feasible. To overcome this challenge and develop isomorphic mapping schemes, it's essential to map the Fock states of photons to qubit states, as discussed in Ref. \cite{Somma_2003}.

In our work, we opted to use Gray code for mapping Fock states to qubit states. The encoding scheme is provided in Table \ref{tabgray}. The gray code utilises the entire Hilbert space, reducing the required number of qubits. Moreover, gray code ensures a Hamming distance of $1$, making it efficient for representing tri-diagonal operators with zeros along the diagonal \citep{Sawaya_2020}.

\begin{table}
    \renewcommand{\arraystretch}{1.5}
    \centering
    \begin{tabular}{|p{1.5cm}|p{3.5cm}|}
    \hline
       Fock State  & Qubit state  \\
    \hline
         $\ket{0}_F$& $\ket{0_{1}0_{2}0_{3}....0_{N_{q}-1}0_{N_{q}}}$\\
          $\ket{1}_F$& $\ket{0_{1}0_{2}0_{3}....0_{N_{q}-1}1_{N_{q}}}$\\
         $\ket{2}_F$& $\ket{0_{1}0_{2}0_{3}....1_{N_{q}-1}1_{N_{q}}}$ \\
         $\ket{3}_F$& $\ket{0_{1}0_{2}0_{3}....1_{N_{q}-1}0_{N_{q}}}$\\
         $\vdots$ & $\vdots$\\
         $\ket{N}_F$& $\ket{1_{1}1_{2}1_{3}....1_{N_{q}-1}1_{N_{q}}}$\\
    \hline
    \end{tabular}
    \vspace{0.5cm}
    \caption{Gray encoding for upto $N$ Fock state. Fock states are mapped to qubit system using $N_{q}$ qubits, where $N=2^{N_{q}}-1$}.
    \label{tabgray}
\end{table}
\subsection{Mapping Bosonic Operators to Pauli operators} \label{sec:Mapping}
Using the gray code representation of Fock states, one can map bosonic creation and annihilation operators to the Pauli operators. Intuitively, this involves applying spin ladder operators on qubits that need to be flipped while the remaining qubits are kept in their present state using the associated projectors. The same logic can be seen applied in Refs. \citep{Di_Matteo_2021,pooja2021} while simulating Deuteron.  

In gray code, for arbitrary Fock states $\ket{n}_F$ and $\ket{n-1}_F$, only the bit at the $i^{th}$ position differs. They are defined as,
\begin{eqnarray*}
    \ket{n}_F \coloneq \ket{g_{1},g_{2},..g_{i},g_{i+1}...g_{N_{q}}},\\
    \ket{n-1}_F \coloneq \ket{g_{1},g_{2},..g'_{i},g_{i+1}...g_{N_{q}}},
    \label{eqn4}
\end{eqnarray*}

where $g_{k} \in \{0,1\}$ is the $k^{th}$ bit in the Gray code. 

The action of $b^{\dagger}$ ($b$) results in the creation (annihilation) of a photon. If photon states are represented in gray code, this action can be simulated by inducing a bit flip at that position. Therefore, we can define $b_{n}^{\dagger}$ which induce $\ket{n-1}_{F} \rightarrow \ket{n}_{F}$ transition as, \begin{equation}
    b^{\dagger}_{n} = \mathcal{P}_{g_{1}}\otimes \mathcal{P}_{g_{2}}\otimes,.. \mathcal{Q}_{g_{i}}\otimes \mathcal{P}_{g_{i+1}}...\mathcal{P}_{g_{N_{q}}}.
    \label{eqn5a}
\end{equation}
Similarly, $b_{n}$ causes $\ket{n}_{F} \rightarrow \ket{n-1}_{F}$ transition and is defined as, 
\begin{equation}
    b_{n} = \mathcal{P}_{g_{1}}\otimes \mathcal{P}_{g_{2}}\otimes,.. \mathcal{Q}_{g'_{i}}\otimes \mathcal{P}_{g_{i+1}}...\mathcal{P}_{g_{N_{q}}}.
    \label{eqn5b}
\end{equation}
Here, $\mathcal{P}_{g_{i}}$ and $\mathcal{Q}_{g_{k}}$ denote the projector and spin ladder operators, respectively and are defined as
\begin{equation}
\begin{aligned}
    \mathcal{P}_{g_{k}}& = 
    \begin{cases}
    \frac{1}{2}\left(\mathbbm{1}+\sigma_{z}\right),&  \text{if } g_{k} = 0\\
    \frac{1}{2}\left(\mathbbm{1}-\sigma_{z}\right),              & 
   \text{if } g_{k} = 1
    \end{cases},  \\ 
    \mathcal{Q}_{g_{k}}& = 
    \begin{cases}
    \frac{1}{2}\left(\sigma_{x}+i\sigma_{y}\right),&\text{if } g_{k} = 0\\
    \frac{1}{2}\left(\sigma_{x}-i\sigma_{y}\right),&\text{if } g_{k} = 1
    \end{cases}. 
\end{aligned}
\label{eqn6}
\end{equation}
We choose the convention of representing of qubit state $|0\rangle = \begin{bmatrix}1\\0\end{bmatrix}$ and $|1\rangle =\begin{bmatrix}0\\1 \end{bmatrix}$. Thus, $\mathcal{Q}_{0}$ act as spin lowering operator and $\mathcal{Q}_{1}$ act as spin raising of operator.

The full creation and annihilation operator can be written in Pauli basis as,
\begin{equation}
    \begin{aligned}
        b_{P}^{\dagger} &= \sum_{n=0}^{N} \sqrt{n+1} b_{n}^{\dagger}, \\
        b_{P} &=\sum_{n=0}^{N} \sqrt{n} b_{n}. 
    \end{aligned}`
    \label{eqn8a}
\end{equation}
Here, $b_{P}\ (b_{P}^{\dagger})$ represents the bosonic creation and annihilation operators in the Pauli basis, and $b_{n}^{\dagger} (b_{n})$ is the tensor product of Pauli operators which is defined as per equations \eqref{eqn5a} and \eqref{eqn5b}.

\section{Simulation of HOM interference}\label {sec:HOMsimulation}
The HOM interference occurs when we input single photons at both input modes. Using two qubits to represent a mode would allow one to encode $3$ fock states, as given in Table \ref{tabgray2}. Simulating HOM interference requires representing each input mode with at least two qubits. 
\begin{table}[h]
    \centering
    \renewcommand{\arraystretch}{1.5}
    \begin{tabular}{|c|c|}
    \hline
       Fock State  & Qubit state  \\
    \hline
         $\ket{0}_F$& $  \ket{00}$ \\

          $\ket{1}_F$& $\ket{01}$ \\
         $\ket{2}_F$& $ \ket{11}$ \\
        $\ket{3}_F$& $ \ket{10}$ \\
    \hline
    \end{tabular}
    \vspace{0.5cm}
    \caption{Gray encoding for upto $3$ Fock state.}
    \label{tabgray2}
\end{table}

In gray code, single photon state $|1\rangle_F$ is represented as $|01\rangle$. Therefore, the circuit should be initialised in the state $|0101\rangle$ before applying the beam splitter unitary. In a two-qubit case, the creation operator induces the following transformations:
\begin{equation}
    \begin{aligned}
        \ket{00} \xrightarrow{b_{0}^{\dagger}=\mathcal{P}_{\small g_1=0}\otimes\mathcal{Q}_{g_2=1}} \ket{01}, \\
        \ket{01} \xrightarrow{b_{1}^{\dagger}=\mathcal{Q}_{g_1=1}\otimes \mathcal{P}_{g_2=1}} \ket{11},\\
        \ket{11} \xrightarrow{b_{2}^{\dagger}=\mathcal{P}_{g_1=1}\otimes \mathcal{Q}_{g_2=0}} \ket{10}. 
    \end{aligned}
\label{eqn7}
\end{equation}

Thus, the bosonic creation and annihilation operator can be represented in Pauli basis as follows,
\begin{eqnarray}
\label{eqn8}
b_{P}^{\dagger}={\mathcal{P}_{0}}\otimes {\mathcal{Q}_{1}}+\sqrt{2} {\mathcal{Q}_{1}}\otimes {\mathcal{P}_{1}} + \sqrt{3} {\mathcal{P}_{1}}\otimes {\mathcal{Q}_{0}},\\
b_{P} ={\mathcal{P}_{0}}\otimes {\mathcal{Q}_{0}}+\sqrt{2} {\mathcal{Q}_{0}}\otimes {\mathcal{P}_{1}} + \sqrt{3} {\mathcal{P}_{1}}\otimes {\mathcal{Q}_{1}}.
\label{eqn9}
\end{eqnarray}

While comparing our mapping scheme with that of Ref. \cite{Di_Matteo_2021}, the primary difference lies in how we realise the flipping of qubits. The qubit flipping was previously realised by employing Pauli-X gate, whereas we use spin ladder operators. A similar mapping of bosonic operators is also observed in Ref. \cite{pooja2021} but they are not using the full creation and annihilation operator that we have given in equation \eqref{eqn8}.


In the mapping scheme of Ref. \cite{Di_Matteo_2021}, we observe that $b^{\dagger}_{n}$ and $b_{n}$ share the same representation. For example, taking $b^{\dagger}_{1} = b_{1} = X\otimes \mathcal{P}_{1}$ following the mapping scheme of Ref. \cite{Di_Matteo_2021}. Consequently, acting $b^{\dagger}_{n}$ twice on the same input state yields the initial state, which contradicts the theory and makes it impossible to simulate HOM interference.

Expanding equation \eqref{eqn8} by substituting equations (\ref{eqn5a}, \ref{eqn5b} and \ref{eqn6} would give the bosonic operators in terms of Pauli operators. Resulting  $b_{P}^{\dagger}$ and $b_{P}$, when plugged into equation \eqref{eqn1}, gives the unitary operator of the beam splitter in the Pauli basis. The unitary operator would comprise terms that exponentiate Pauli strings, which can be easily decomposed into standard gates using trotterization and canonical quantum circuit creation methods. Figure \ref{circuit1} depicts the structure of the quantum circuit that simulates the HOM interference. Further discussion on the exact construction of the quantum circuit from the beam splitter unitary can be found in the \textit{supplementary material 1}.

\begin{figure*}[t!]
\centering
\includegraphics[width=\textwidth]{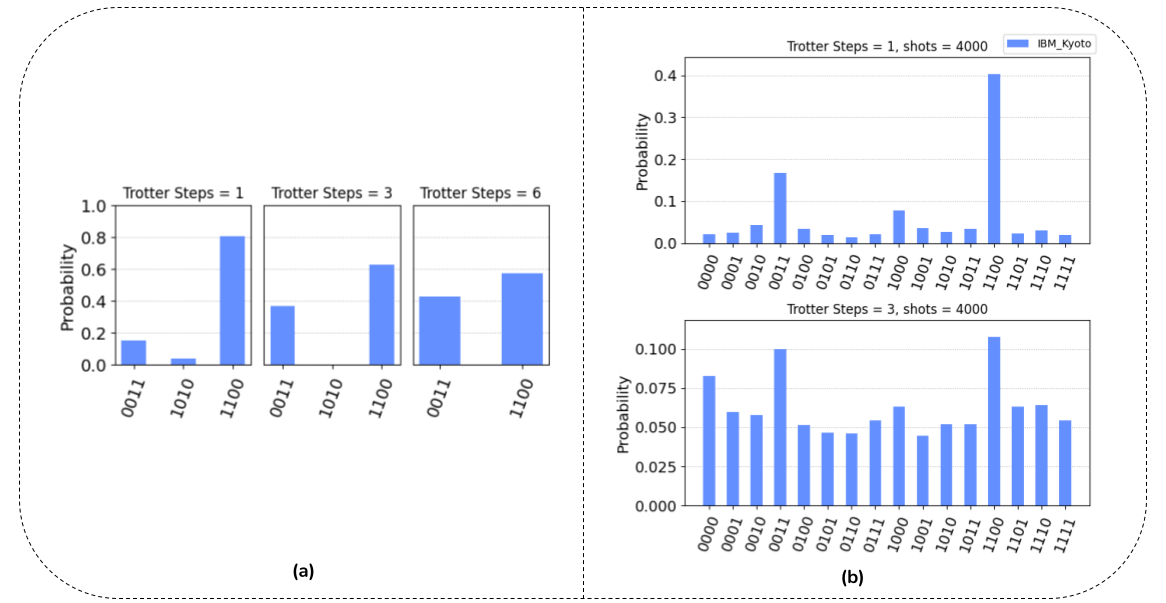}
\caption{Simulation results of HOM interference experiment. (a) Shows the result of the simulation performed in Qiskit AerSimulator. (b) Shows the result of a simulation performed in IBM Kyoto.}
\label{figresult}
\end{figure*}
\subsection{Circuit Size Reduction}
\label{sec:circuitreduction}

Simulating HOM interference on actual backends posed significant challenges due to the extensive circuit depth and a substantial number of \textit{CX} gates. To illustrate the circuit's size, for a single Trotter step ($\text{trotter step} = 1$), the circuit depth reached $178$ layers, necessitating $128$ \textit{CX} operations. Observing HOM interference under these conditions proved challenging on current hardware. Consequently, we employed specific circuit size reduction strategies tailored to the problem, enabling us to simulate the experiment on existing quantum computer successfully.

During HOM interference, the total number of photons will be conserved. Given the initial state of two photons, the output state must also contain two photons, as expressed by the equation,
\begin{equation*}
    \hat{U}_{B} |0101\rangle = \frac{1}{\sqrt{2}}\left[|1100\rangle + |0011\rangle\right].
\end{equation*}
The dynamics of the system only span across the basis $\{|00\rangle,|01\rangle,|11\rangle\}$ rendering the basis $|10\rangle$ redundant. Consequently, terms in $b_{P}^{\dagger}$ ($b_{P}$) that cause transitions to (from) the state $|10\rangle$ can be eliminated. This allows us to rewrite equations \eqref{eqn8} and \eqref{eqn9} as,
\begin{eqnarray*}
    b_{P}^{\dagger}={\mathcal{P}_{0}}\otimes {\mathcal{Q}_{1}}+\sqrt{2} {\mathcal{Q}_{1}}\otimes {\mathcal{P}_{1}},\\
    b_{P} ={\mathcal{P}_{0}}\otimes {\mathcal{Q}_{0}}+\sqrt{2} {\mathcal{Q}_{0}}\otimes {\mathcal{P}_{1}}.
\end{eqnarray*}

One might argue that since each mode is initialised to the state $|01\rangle$, terms like $P_{0}\otimes Q_{1}$ and $\sqrt{2}Q_{0}\otimes P_{1}$ do not affect the state and can be removed. However, removing these terms renders $\hat{U}_{B}$ non-unitary, making it challenging to simulate the effect. Nevertheless, further circuit size reduction is possible after constructing $\hat{U}_{B}$ by identifying terms that do not impact the input states. Refer to the \textit{supplementary material 2} for more details. 

\section{Results}\label{results}
\textbf{On simulator} we performed simulation of HOM interference using Qiskit's \textit{AerSimulator} \citep{Qiskit}, and the outcomes are presented in Figure \ref{figresult}. The simulation was performed for various trotter steps while maintaining the number of shots to $10,000$ for each case. The results indicate a suppression of occurrences of the state $|0101\rangle$ (representing one photon across each output mode) with increasing trotter steps. Additionally, it is evident that the probabilities of states $|1100\rangle$ and $|0011\rangle$ become more pronounced, approaching equal probability values as trotter step increases.

\textbf{On hardware}, the simulation was performed using Qiskit's Runtime Sampler for two trotter steps while maintaining the number of shots to $4000$ in each case. Simulations were performed using a quantum circuit whose size was reduced as described in section \ref{sec:circuitreduction}. The results are presented in the figure \ref{figresult}. Due to errors in quantum computation, states other than the expected states have non-zero probabilities. Nevertheless, the expected states $|1100\rangle$ and $|0011\rangle$ show higher probability than others. As the number of trotter steps increases, gate count and circuit depth increase, leading to the equilibration of probabilities of all possible states.

\section{Discussion and Outlook}\label{conclusion}

In this paper, we have detailed the construction of a quantum circuit that emulates a beam splitter using gray encoding. To construct the circuit, we used the quantum description of the beam splitter, sometimes called quantum beamsplitter formalism. We validated our approach by simulating the famous HOM interference experiment in a quantum computer. Even though we only demonstrated HOM interference, the construction of the circuit is much more general and can be used to study various phenomena associated with beam splitters.

It is worth noting that DQS of the beam splitter has been previously discussed in \cite{Encinar_2021}. However, they adopted a unary encoding while we used gray code to encode the states. Our interest in gray encoding was fuelled by the results of \cite{Sawaya_2020} showing the superiority of gray code over unary, requiring fewer qubits and gate operations. Given the change in state encoding, appropriate changes are necessary for mapping bosonic operators to Pauli operators. Additionally, \cite{Encinar_2021} did not explicitly address the simulation of HOM interference.

While working on this paper, we came across a recent work by \cite{chin2022quantum} that exploits boson-fermion correspondence relations to propose a novel boson-to-qubit mapping scheme. They transform bosonic states into fermionic states with internal degrees of freedom, which are then mapped to qubit states using the well-known Jordan-Wigner transformation. They demonstrate the effectiveness of this scheme by simulating the HOM interference experiment. Furthermore, they could simulate the HOM dip by introducing distinguishability within qubit states using additional internal degrees of freedom. 

In comparison, our mapping scheme inherently treats photons as indistinguishable since, at the core, it is an integer-to-bit mapping. We are only concerned with the number of photons in each mode and treat other features of photons, such as polarisation states, as identical. To introduce distinguishability among photons, exploring methods for incorporating additional degrees of freedom into our simulation is essential. 

Using integer-to-bit mapping to encode additional degrees of freedom may be costly since the qubit requirement is significant. For instance, if we want to encode an arbitrary polarisation state $\ket{\beta} = \frac{1}{\sqrt{2}} (a \ket{H} + b \ket{V})$, using integer-to-bit mapping requires truncation of states as we do for Fock states. An alternative and qubit-efficient approach to encode additional degrees of freedom is amplitude encoding. But irrespective of the encoding scheme chosen, one needs to make appropriate changes to $\hat{U}_{B}$ to incorporate additional degrees of freedom.

Finding qubit-efficient ways to encode additional degrees of freedom is a promising and intriguing avenue of research. Efficient encoding techniques play a pivotal role in advancing the field of digital quantum simulation, offering the potential to simulate complex quantum systems with fewer qubits and improved computational resources. Such developments could significantly enhance the feasibility and scalability of DQS across various applications, marking a crucial step forward in harnessing the power of quantum computing for realistic and impactful simulations.

\section{Acknowledgements}
We would like to acknowledge all who have contributed to this work through their meaningful discussions and suggestions. This work is a part of the QuRNG project of Qulabs Software India. 
 
\bibliographystyle{unsrt}
\bibliography{references}  

\twocolumn[
  \begin{@twocolumnfalse}
    \begin{center}
        \begin{dmath}
        \displaystyle \hat{b}^{\dagger}_{P}\hat{a}_{P}+\hat{b}_{P}\hat{a}^{\dagger}_{P}= - \frac{1 \left({I^{b}}\otimes {\sigma_{x}^{b}}\otimes {\sigma_{z}^{a}}\otimes {\sigma_{x}^{a}} + {I^{b}}\otimes {\sigma_{y}^{b}}\otimes {\sigma_{z}^{a}}\otimes {\sigma_{y}^{a}} + {\sigma_{z}^{b}}\otimes {\sigma_{x}^{b}}\otimes {I^{a}}\otimes {\sigma_{x}^{a}} + {\sigma_{z}^{b}}\otimes {\sigma_{y}^{b}}\otimes {I^{a}}\otimes {\sigma_{y}^{a}}\right)}{4} + \frac{\left(2 - \sqrt{3}\right) \left({I^{b}}\otimes {\sigma_{y}^{b}}\otimes {I^{a}}\otimes {\sigma_{y}^{a}} + {\sigma_{z}^{b}}\otimes {\sigma_{x}^{b}}\otimes {\sigma_{z}^{a}}\otimes {\sigma_{x}^{a}}\right)}{4} + \frac{\left(2 + \sqrt{3}\right) \left({I^{b}}\otimes {\sigma_{x}^{b}}\otimes {I^{a}}\otimes {\sigma_{x}^{a}} + {\sigma_{z}^{b}}\otimes {\sigma_{y}^{b}}\otimes {\sigma_{z}^{a}}\otimes {\sigma_{y}^{a}}\right)}{4}
        - \frac{1 \left({I^{b}}\otimes {\sigma_{y}^{b}}\otimes {\sigma_{y}^{a}}\otimes {I^{a}} - {I^{b}}\otimes {\sigma_{y}^{b}}\otimes {\sigma_{y}^{a}}\otimes {\sigma_{z}^{a}} + {\sigma_{z}^{b}}\otimes {\sigma_{x}^{b}}\otimes {\sigma_{x}^{a}}\otimes {I^{a}} - {\sigma_{z}^{b}}\otimes {\sigma_{x}^{b}}\otimes {\sigma_{x}^{a}}\otimes {\sigma_{z}^{a}}\right)}{4}+ \frac{\left(2 - \sqrt{6}\right) \left({I^{b}}\otimes {\sigma_{x}^{b}}\otimes {\sigma_{x}^{a}}\otimes {I^{a}} - {I^{b}}\otimes {\sigma_{x}^{b}}\otimes {\sigma_{x}^{a}}\otimes {\sigma_{z}^{a}} + {\sigma_{z}^{b}}\otimes {\sigma_{y}^{b}}\otimes {\sigma_{y}^{a}}\otimes {I^{a}} - {\sigma_{z}^{b}}\otimes {\sigma_{y}^{b}}\otimes {\sigma_{y}^{a}}\otimes {\sigma_{z}^{a}}\right)}{8}
        \\- \frac{1 \left({\sigma_{x}^{b}}\otimes {I^{b}}\otimes {\sigma_{z}^{a}}\otimes {\sigma_{x}^{a}} - {\sigma_{x}^{b}}\otimes {\sigma_{z}^{b}}\otimes {\sigma_{z}^{a}}\otimes {\sigma_{z}^{a}} + {\sigma_{y}^{b}}\otimes {I^{b}}\otimes {I^{a}}\otimes {\sigma_{y}^{a}} - {\sigma_{y}^{b}}\otimes {\sigma_{z}^{b}}\otimes {I^{a}}\otimes {\sigma_{y}^{a}}\right)}{4} + \frac{\left(2 - \sqrt{6}\right) \left({\sigma_{x}^{b}}\otimes {I^{b}}\otimes {I^{a}}\otimes {\sigma_{x}^{a}} - {\sigma_{x}^{b}}\otimes {\sigma_{z}^{b}}\otimes {I^{a}}\otimes {\sigma_{x}^{a}} + {\sigma_{y}^{b}}\otimes {I^{b}}\otimes {\sigma_{z}^{b}}\otimes {\sigma_{y}^{b}} - {\sigma_{y}^{b}}\otimes {\sigma_{y}^{b}}\otimes {\sigma_{z}^{a}}\otimes {\sigma_{y}^{a}}\right)}{8}
        \\+ \frac{1 \left({\sigma_{x}^{b}}\otimes {I^{b}}\otimes {\sigma_{x}^{a}}\otimes {I^{a}} - {\sigma_{x}^{b}}\otimes {I^{b}}\otimes {\sigma_{x}^{a}}\otimes {\sigma_{z}^{a}} - {\sigma_{x}^{b}}\otimes {\sigma_{z}^{b}}\otimes {\sigma_{x}^{a}}\otimes {I^{a}} + {\sigma_{x}^{b}}\otimes {\sigma_{z}^{b}}\otimes {\sigma_{x}^{a}}\otimes {\sigma_{z}^{a}} + {\sigma_{y}^{b}}\otimes {I^{b}}\otimes {\sigma_{y}^{a}}\otimes {I^{a}}\right)}{4} - \frac{1\left({\sigma_{y}^{b}}\otimes {I^{b}}\otimes {\sigma_{y}^{a}}\otimes {\sigma_{z}^{a}} - {\sigma_{y}^{b}}\otimes {\sigma_{z}^{b}}\otimes {\sigma_{y}^{a}}\otimes {I^{a}} + {\sigma_{y}^{b}}\otimes {\sigma_{z}^{b}}\otimes {\sigma_{y}^{a}}\otimes {\sigma_{y}^{b}}\right)}{4}
        \label{supeqn1}
        \end{dmath}
    \end{center}
\end{@twocolumnfalse}
]

\section*{}
\begin{center}
    \textbf{\large 1. Supplementary Material: Construction of Quantum Circuit}
    \label{supplementary1}
\end{center}

\vspace{1cm}
The unitary of the beam splitter is given by equation \ref{eqn1}. As we discussed in the section \ref{sec:Mapping}, using the Gray code to represent the Fock states, one can map the bosonic creation and annihilation operators to Pauli operators as per equation \eqref{eqn8} and \eqref{eqn9}. In such a case, the two-site bosonic interaction operator can be written in Pauli basis as given in \eqref{supeqn1}.

The beam splitter unitary $\hat{U}_{B}$ in terms of Pauli operators by substituting equation \eqref{supeqn1} back to equation \eqref{eqn1}. The unitary will be on composed of terms that exponentiate Pauli strings and could be easily decomposed into a quantum circuit. The canonical quantum circuits used to exponentiate Pauli strings on a universal quantum computer are given in the table \ref{tab3}.

\begin{center}
\begin{table}
\centering
\begin{tabular}{|c|c|}
\hline
\textbf{Term} &\textbf{Circuit} \\
 \hline
 $\exp(-i\theta \sigma_{x})$ & Figure \ref{circuit 1}\\
 $\exp(-i\theta\sigma_{x}\otimes\sigma_{y})$ & Figure \ref{circuit2}\\
 $\exp(-i\theta\sigma_{x}\otimes\sigma_{z}\otimes\sigma_{y})$ & Figure \ref{circuit3}\\
 $\exp(-i\theta\sigma_{x}\otimes I\otimes\sigma_{z}\otimes\sigma_{y})$ & Figure \ref{circuit4}\\
 \hline
\end{tabular}
\vspace{0.5cm}
\caption{Canonical quantum circuits used to exponentiate Pauli strings}
\label{tab3}
\end{table}
\end{center}

\begin{center}
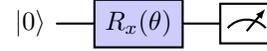

\tikzset{
operator/.append style={fill=blue!20},
my label/.append style={above right,xshift=1cm},
phase label/.append style={label position=above}
} 
\begin{quantikz}
    \lstick{\ket{0}}&\gate{R_{x}(\theta)}&\meter{}\\
\end{quantikz}
\captionof{figure}{Circuit for 1st term in Table \eqref{tab3}}
\label{circuit 1}
\end{center}
\begin{center}
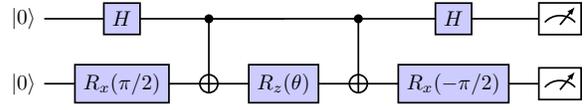

\tikzset{
operator/.append style={fill=blue!20},
my label/.append style={above right,xshift=1cm},
phase label/.append style={label position=above}
} 
\begin{tikzpicture}
\node[scale=0.8]{
    \begin{quantikz}
        \lstick{\ket{0}}&\gate{H}&\ctrl{1}&\qw&\ctrl{1}&\gate{H}&\meter{}\\

        \lstick{\ket{0}}&\gate{R_{x}(\pi/2)}&\targ{}&\gate{R_{z}(\theta)}&\targ{}&\gate{R_{x}(-\pi/2)}&\meter{}\\
    \end{quantikz}
};
\end{tikzpicture}
\captionof{figure}{Circuit for 2nd term in Table \eqref{tab3}}
\label{circuit2}
\end{center}

\begin{center}
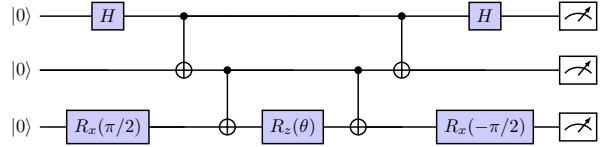

\tikzset{
operator/.append style={fill=blue!20},
my label/.append style={above right,xshift=1cm},
phase label/.append style={label position=above}
} 
\begin{tikzpicture}
\node[scale=0.7]{
    \begin{quantikz}
        \lstick{\ket{0}}&\gate{H}&\ctrl{1}&\qw&\qw&\qw&\ctrl{1}&\gate{H}&\meter{}\\
        
        \lstick{\ket{0}}&\qw&\targ{}&\ctrl{1}&\qw&\ctrl{1}&\targ{}&\qw & \meter{}\\

        \lstick{\ket{0}}&\gate{R_{x}(\pi/2)}&\qw&\targ{}&\gate{R_{z}(\theta)}&\targ{}&\qw&\gate{R_{x}(-\pi/2)}&\meter{}\\
    \end{quantikz}
};
\end{tikzpicture}
\captionof{figure}{Circuit for 3rd term in Table \eqref{tab3}}
\label{circuit3}
\end{center}

\begin{center}
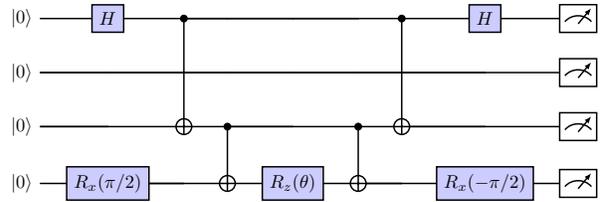

\tikzset{
operator/.append style={fill=blue!20},
my label/.append style={above right,xshift=1cm},
phase label/.append style={label position=above}
} 
\begin{tikzpicture}
\node[scale=0.7]{
    \begin{quantikz}
        \lstick{\ket{0}}&\gate{H}&\ctrl{2}&\qw&\qw&\qw&\ctrl{2}&\gate{H}&\meter{}\\
        
        \lstick{\ket{0}}&\qw&\qw&\qw&\qw&\qw&\qw&\qw & \meter{}\\
        
        \lstick{\ket{0}}&\qw&\targ{}&\ctrl{1}&\qw&\ctrl{1}&\targ{}&\qw & \meter{}\\

        \lstick{\ket{0}}&\gate{R_{x}(\pi/2)}&\qw&\targ{}&\gate{R_{z}(\theta)}&\targ{}&\qw&\gate{R_{x}(-\pi/2)}&\meter{}\\
    \end{quantikz}
};
\end{tikzpicture}
\captionof{figure}{Circuit for 4th term in Table \eqref{tab3}}
\label{circuit4}
\end{center}
\newpage

\twocolumn[
  \begin{@twocolumnfalse}
  
\section*{}\label{supplementary2}
\begin{center}
    \textbf{\large 2. Supplementary Material: Circuit Size Reduction}
\end{center}
As discussed in the section \ref{sec:circuitreduction}, in the context of the HOM interference, the basis $|10\rangle$ is redundant. This leads to a simplified representation of $\hat{b}^{\dagger}{P}$ and $\hat{b}{P}$ as provided below.

\begin{eqnarray}
    \label{suppleqn3}
    b_{P}^{\dagger}={\mathcal{P}_{0}}\otimes {\mathcal{Q}_{1}}+\sqrt{2} {\mathcal{Q}_{1}}\otimes {\mathcal{P}_{1}}\\
    b_{P} ={\mathcal{P}_{0}}\otimes {\mathcal{Q}_{0}}+\sqrt{2} {\mathcal{Q}_{0}}\otimes {\mathcal{P}_{1}}
    \label{suppleqn2}
\end{eqnarray}

It necessarily blocks any transition to or from the $|10\rangle$, which is valid since the photon number is always conserved. Furthermore, one can also argue to remove the terms $\mathcal{P}_{0}\otimes \mathcal{Q}_{1}$ and $\sqrt{2}\mathcal{Q}_{0}\otimes\mathcal{P}_{1}$ since:
\begin{eqnarray*}
    \mathcal{P}_{0}\otimes \mathcal{Q}_{1} |01\rangle &=& 0\\
    \sqrt{2}\mathcal{Q}_{0}\otimes\mathcal{P}_{1} |01\rangle &=& 0
\end{eqnarray*}
While expressing $\hat{b}^{\dagger}{P}$ as $\sqrt{2} \mathcal{Q}{1}\otimes\mathcal{P}{1}$ and $\hat{b}{P}$ as $\mathcal{P}{0}\otimes\mathcal{Q}{0}$ might seem appealing for circuit simplification, it introduces a challenge. he issue lies in the construction of $\hat{U}{B}$, which will no longer be unitary with this representation. One can easily verify this by constructing the two-site bosonic interaction operator and substituting it into the expression of $\hat{U}_{B}$. There would be terms of the form $\exp(\hat{M})$, which will be unitary only if $\hat{M}$ is skew hermitian. However, in this case, $\hat{M}$ turns out to be Hermitian since it is a tensor product of Pauli operators.\\

Nonetheless, one can further reduce the circuit size by identifying the terms which do not affect the computation and removing them during circuit construction. From the equations \eqref{suppleqn3} and \eqref{suppleqn2} we can derive the two-site bosonic interaction operator as:

\begin{dmath}
    \begin{aligned}
        \hat{b}_{P}^{\dagger}\hat{a}_{P}+ \hat{b}_{P}\hat{a}_{P}^{\dagger} = & (\mathcal{P}_{0}\otimes \mathcal{Q}_{1}\otimes \mathcal{P}_{0}\otimes \mathcal{Q}_{0}+ h.c) +\\
        &\sqrt{2}(\mathcal{P}_{0}\otimes \mathcal{Q}_{1}\otimes \mathcal{Q}_{0}\otimes \mathcal{P}_{1}+h.c)+\\
        &\sqrt{2}(\mathcal{Q}_{1}\otimes \mathcal{P}_{1}\otimes \mathcal{P}_{0}\otimes \mathcal{Q}_{0}+h.c)+\\
        &2(\mathcal{Q}_{1}\otimes \mathcal{P}_{1}\otimes \mathcal{Q}_{0}\otimes \mathcal{P}_{1}+h.c)\\
     \label{suppleqn5}
     \end{aligned}
\end{dmath}

We observed that the first and last terms of equation \eqref{suppleqn5} do not influence the HOM interference experiment. Thus, they can be removed from the expression of the two-site bosonic interaction term without affecting the dynamics. Consequently, the two-site bosonic interaction operator effectively becomes:
\begin{dmath}
    \begin{aligned}
        \hat{b}_{P}^{\dagger}\hat{a}_{P}+ \hat{b}_{P}\hat{a}_{P}^{\dagger} \approx \sqrt{2}\{(\mathcal{P}_{0}\otimes \mathcal{Q}_{1}\otimes \mathcal{Q}_{0}\otimes \mathcal{P}_{1}+h.c)+(\mathcal{Q}_{1}\otimes \mathcal{P}_{1}\otimes \mathcal{P}_{0}\otimes \mathcal{Q}_{0}+h.c)\}\\
    \end{aligned}
    \label{suppleqn6}
\end{dmath}

By constructing $\hat{U}_{B}$ from the two-site bosonic interaction given in \eqref{suppleqn6} and following the quantum circuit creation method detailed in \textbf{Supplementary Section 1}, one can build a reduced quantum circuit that demonstrates the HOM interference effect.
\end{@twocolumnfalse}]

\end{document}